\documentclass{comex}
\usepackage[T1]{fontenc}
\usepackage{stix}

\newtheorem{theorem}{Theorem}
\newtheorem{definition}[theorem]{Definition}

\title{Introducing the Perception-Distortion Tradeoff into the Rate-Distortion Theory of General Information Sources\thanks{%
The result in this letter will be presented at
the 41st Symposium on Information Theory and its Applications,
Fukushima, Japan, 18--21 December 2018. The author was given the
right to upload this manuscript to arxiv.org
as stated at the footnote 2 in page 13 of
\url{http://www.ieice.org/eng/copyright/files/copyright.pdf}}}
\author{Ryutaroh Matsumoto$^{\rm 1a)}$}
\affiliate{$^{1}$
  Department of Information and Communication Engineering,
  Nagoya University\\
Furo-cho, Chitane-ku, Nagoya, Aichi 464-8603, Japan}

\email{{\rm a)} ryutaroh.matsumoto@nagoya-u.jp}
\received{2018}{8}{2}
\accepted{2018}{8}{23}
\vol{7}
\no{12}

\begin{document}
\sloppy
\vol{7}
\no{12}
\maketitle

\begin{abstract}
Blau and Michaeli recently introduced a novel concept
  for inverse problems of signal processing, that is,
  the perception-distortion tradeoff.
  We introduce their tradeoff into the rate distortion theory
  of lossy source coding in information theory,
  and clarify the tradeoff among information rate, distortion
  and perception for general information sources.
\end{abstract}
\begin{keywords}
perception-distortion tradeoff, rate-distortion theory, data compression
\end{keywords}
\begin{classification}
Fundamental theories for communications
\end{classification}

%\bibliographystyle{comex} % Bibtex style file for ComEX
%\bibliography{rd} % Sample bibtex source file

\section{Introduction}
An inverse problem of signal processing
is to reconstruct the original information from its degraded
version. It is not limited to image processing, but it often
arises in the image processing.
When a natural image is reconstructed, the reconstructed
image sometimes does not look natural while it is close to
the original image by a reasonable metric, for example
mean squared error.
When the reconstructed information is close to the original,
it is often believed that it should also look natural.

Blau and Michaeli \cite{cvpr2018}
questioned this unproven belief.
In their research \cite{cvpr2018},
they mathematically formulated the \emph{naturalness}
of the reconstructed information by a
distance between the probability distributions of
the reconstructed information and the original information.
The reasoning behind this is that the perceptional quality of
a reconstruction method is often evaluated by
how often a human observer can distinguish an output of
the reconstruction method from natural ones.
Such a subjective evaluation can mathematically be
modeled as a hypothesis testing \cite{cvpr2018}.
A reconstructed image is more easily distinguished
as the variational distance
$\sigma(P_R$, $P_N)$ increases \cite{cvpr2018},
where $P_R$ is the probability distribution of
the reconstructed information and
$P_N$ is that  of
the natural one.
They regard the perceptional quality of reconstruction as
a distance between $P_R$ and $P_N$.
The distance between the reconstructed information
and the original information is conventionally called as  distortion.
They discovered that there exists a tradeoff
between perceptional quality and distortion, and
named it as the \emph{perception-distortion tradeoff}.

Claude Shannon \cite[Chapter 5]{hanbook} initiated the rate-distortion
theory in 1950's.
It clarifies the tradeoff between
information rate and distortion in the lossy source coding
(lossy data compression).
The rate-distortion theory has served as a theoretical
foundation of image coding for past several decades, as
drawing a rate-distortion curve is a common practice in
research articles of image coding.
Since distortion and perceptional quality are now considered
two different things,
it is natural to consider a tradeoff among
information rate, distortion and perceptional quality.
Blau and Michaeli \cite{cvpr2018}
briefly mentioned the rate-distortion theory,
but they did not clarify the tradeoff among the three.

The purpose of this letter is to mathematically
define the tradeoff for general information sources,
and to express the tradeoff in terms of
information spectral quantities introduced by Han and Verd\'u
\cite{hanbook}.
It should be noted that
the tradeoff among the three quantities can be regarded as a
combination of lossy source coding problem
\cite[Chapter 5]{hanbook} and random number generation
problem \cite[Chapter 2]{hanbook},
both of which will be used to derive the tradeoff.

Since the length limitation is strict in this journal,
citations to the original papers are replaced by
those to the textbook \cite{hanbook}, and
the mathematical proof is a bit compressed.
The author begs readers' kind understanding.
The base of $\log$ is an arbitrarily fixed real number $>1$
unless otherwise stated.

\section{Preliminaries}
The following definitions are borrowed from Han's textbook
\cite{hanbook}.
Let
\[
\mathbf{X} = \left\{ X^n = (X_1^{(n)}, \ldots, X_n^{(n)} ) \right\}_{n=1}^\infty
\]
be a general information source, where the alphabet of the random variable $X^n$
is the $n$-th Cartesian product $\mathcal{X}^n$ of some finite alphabet
$\mathcal{X}$.
For a sequence of real-valued random variables $Z_1$, $Z_2$, \ldots
we define
\[
\textrm{p-}\limsup_{n\rightarrow \infty} Z_n =\inf\left\{
\alpha \mid \lim_{n\rightarrow\infty} \mathrm{Pr}[Z_n > \alpha] = 0 \right\}.
\]
For two general information sources $\mathbf{X}$ and $\mathbf{Y}$ we define
\[
\overline{I}(\mathbf{X}; \mathbf{Y}) = \textrm{p-}\limsup_{n\rightarrow \infty}
\frac{1}{n} \log \frac{P_{X^nY^n}(X^n, Y^n)}{P_{X^n}(X^n)P_{Y^n}(Y^n)},
\]
and
\[
F_\mathbf{X}(R) = \limsup_{n\rightarrow \infty} \mathrm{Pr}\left[
  \frac{1}{n} \log \frac{1}{P_{X^n}(X^n)} \geq R \right].
\]

For two distributions $P$ and $Q$ on an alphabet $\mathcal{X}$,
we define the variational distance $\sigma(P,Q)$ as $\sum_{x\in \mathcal{X}}
|P(x)-Q(x)|/2$. In the rate-distortion theory, we usually
assume a reconstruction alphabet different from a source alphabet.
In order to consider the distribution similarity of reconstruction,
in this letter we assume $\mathcal{X}^n$ as both source and
reconstruction alphabets.

An encoder of length $n$ is a mapping $f_n: \mathcal{X}^n \rightarrow
\{1$, \ldots, $M_n\}$, and
the corresponding decoder of length $n$
is a mapping
$g_n: \{1$, \ldots, $M_n\} \rightarrow \mathcal{X}^n$.
$\delta_n: \mathcal{X}^n \times \mathcal{X}^n
\rightarrow [0,\infty)$ is a general distortion function
  with the assumption
  $\delta_n(x^n, x^n)=0$ for all $n$ and $x^n \in \mathcal{X}^n$.

\begin{definition}
A triple $(R,D,S)$ is said to be achievable
if there exists a sequence of encoder and decoder
$(f_n$, $g_n)$ such that
\begin{eqnarray}
  \limsup_{n\rightarrow \infty} \frac{\log M_n}{n} & \leq & R,\label{eq3}\\
  \textrm{p-}\limsup_{n\rightarrow \infty} \frac{1}{n}\delta_n(X^n, g_n(f_n(X^n))) & \leq & D,\label{eq4}\\
  \limsup_{n\rightarrow \infty} \sigma(P_{g_n(f_n(X^n))}, P_{X^n}) & \leq & S.\label{eq5}
\end{eqnarray}
Define the function $R(D,S)$ by
\[
R(D,S) = \inf\{ R \mid (R,D,S) \mbox{ is achievable }\}.
\]
\end{definition}

\begin{theorem}\label{thm1}
  \[
  R(D,S) = \max\left\{ \inf_{\mathbf{Y}} \overline{I}(\mathbf{X}; \mathbf{Y}),
  \inf\{R \mid F_{\mathbf{X}}(R) \leq S\} \right\}
  \]
  where the infimum is taken with respect to all
  general information sources $\mathbf{Y}$ satisfying
  \begin{equation}
    \textrm{p-}\limsup_{n\rightarrow \infty} \frac{1}{n}\delta_n(X^n, Y^n)  \leq  D.\label{eq1}
  \end{equation}
\end{theorem}

\noindent\textbf{Proof:}
Let a pair of encoder $f_n$ and decoder $g_n$ satisfies
Eqs.\ (\ref{eq3})--(\ref{eq5}).
Then by \cite[Theorem 5.4.1]{hanbook} we have
\begin{equation}
  R \geq \inf_{\mathbf{Y}} \overline{I}(\mathbf{X}; \mathbf{Y}),\label{eq30}
\end{equation}
where $\mathbf{Y}$ satisfies Eq.\ (\ref{eq1}).
On the other hand,
the decoder $g_n$ can be viewed as a random number generator
to $X^n$ from the alphabet $\{1$, \ldots, $M_n\}$.
By \cite[Converse part of the proof of Theorem 2.4.1]{hanbook} we
have
\begin{equation}
  R \geq \inf\{R \mid F_{\mathbf{X}}(R) \leq S\}. \label{eq31}
\end{equation}
This complete the converse part of the proof.

We start the direct part of the proof.
Assume that  a triple $(R,D,S)$ satisfys Eqs. (\ref{eq30}) and (\ref{eq31}).
Let $M_n$ satisfy
\begin{equation}
  \limsup_{n\rightarrow\infty} \frac{1}{n}\log M_n  \leq  R. \label{eq11}
\end{equation}
Let $f_n^{(1)}$ and $g_n^{(1)}$ be an encoder and a decoder
constructed in \cite[Lemma 1.3.1]{hanbook}
with codebook $\{1$, \ldots, $M_n\}$.
Let $f_n^{(2)}$ and $g_n^{(2)}$ be an encoder and a decoder
constructed in \cite[Theorem 5.4.1]{hanbook}
with codebook $\{M_n+1$, \ldots, $2M_n\}$.
Assume that we have a source sequence $x^n \in \mathcal{X}^n$.
If $- \log P_{X^n}(x^n) < \log M_n$ then
let $f_n^{(1)}(x^n) \in \{1$, \ldots, $M_n\}$ be the codeword.
If $- \log P_{X^n}(x^n) \geq \log M_n$ then
let $f_n^{(2)}(x^n) \in \{M_n+1$, \ldots, $2M_n\}$ be the codeword.
Let $f_n$ be the above encoding process.
At the receiver of a codeword $1 \leq m \leq 2M_n$,
if $m \leq M_n$ then decode $m$ by $g_n^{(1)}$,
otherwise decode $m$ by by $g_n^{(2)}$.
Let the above decoding process as $g_n$.

If $f_n^{(1)}$ and $g_n^{(1)}$  are used
then the source sequence $x^n$
is reconstructed by a receiver without error by \cite[Lemma 1.3.1]{hanbook}
and we have $\delta_n(x^n, g_n(f_n(x^n)))=0$.
The probability $\epsilon_n$ of $f_n^{(1)}$ and $g_n^{(1)}$ \emph{not}
being used is
\[
  \epsilon_n \leq \mathrm{Pr}\left[ \frac{1}{n} \log
    \frac{1}{P_{X^n}(X^n)} \geq \frac{1}{n} \log M_n \right].
\]
Combined with the assumption $R \geq \inf\{R \mid F_{\mathbf{X}}(R) \leq S\}$ and Eq.\ (\ref{eq11})
\[
  \limsup_{n\rightarrow\infty} \epsilon_n  \leq  S,
\]
which implies Eq.\ (\ref{eq5}).

On the other hand, $f_n^{(2)}$ and $g_n^{(2)}$ satisfy
Eq.\ (\ref{eq4}), so the combined encoder $f_n$ and $g_n$ also
satisfies Eq.\ (\ref{eq4}).
The information rate of $f_n$ is at most $R+ \frac{\log 2}{n}$,
which implies that Eq.\ (\ref{eq3}) holds with the constructed
$f_n$ and $g_n$. This completes the direct part of the proof.
\rule{1ex}{1ex}

\section{Example with a mixed information source}
A typical example of non-ergordic general information source
is a mixed information source \cite[Section 1.4]{hanbook}.
Since Theorem \ref{thm1} is a bit abstract,
we explicitly compute $R(D,S)$ for a mixed information source.
Let $\mathcal{X} =\{0$, $1\}$,
and $\delta_n(x^n$, $y^n)$ be the Hamming distance
between $x^n$, $y^n \in \mathcal{X}^n$.
Consider two distributions $P$ and $Q$ on  $\mathcal{X}$
defined by
\begin{eqnarray*}
  && P(0) = 1/2, P(1)=1/2,\\
  && Q(0) = 1/4, Q(1)=3/4.
\end{eqnarray*}
For $x^n = (x_1$, \ldots, $x_n)$, in our mixed information source
we have
\[
\mathrm{Pr}[X^n = x^n] = \frac{1}{2} \prod_{i=1}^n P(x_i) + \frac{1}{2} \prod_{i=1}^n Q(x_i).
\]
By \cite[Theorem 5.8.1, Example 5.8.1 and Theorem 5.10.1]{hanbook}
we see that
\[
R \geq \inf_{\mathbf{Y}: \textrm{\scriptsize Eq.\ (\ref{eq1}) holds}} \overline{I}(\mathbf{X}; \mathbf{Y})
\]
if and only if
\begin{equation}
  R \geq h(1/2) - h(D), \label{eq21}
\end{equation}
where $h(u)$ is the binary entropy function $-u \log u - (1-u)\log (1-u)$.

On the other hand, by \cite[Example 1.6.1]{hanbook},
we have
\[
F_{\mathbf{X}}(R) = \left\{
\begin{array}{cc}
  1 & \textrm{ if } R < h(1/4),\\
  1/2 & \textrm {if } h(1/4) \leq R < 1, \\
  0 & \textrm {if } 1 \leq R.
\end{array}
\right.
\]
By the above formulas and assuming $\log = \log_2$, we can see
\[
R(D,S) = \left\{
\begin{array}{cc}
  1 & \textrm{ if } S=0,\\
  \max\{ h(1/4), 1-h(D)\} & \textrm {if } 0< S \leq 1/2, \\
  1-h(D) & \textrm {if } 1/2 < S.
\end{array}
\right.
\]

\section*{Acknowledgments}
The author would like to thank Dr.\ Tetsunao Matsuta for
the helpful discussions.

\end{document}